# Bond-dependent anisotropy and magnon breakdown in cobalt Kitaev triangular antiferromagnet

Chaebin Kim[1,2,$], Sujin Kim[3,$], Pyeongjae Park[1,2], Taehun Kim[1,2], Jaehong Jeong[1,2], Seiko Ohira-Kawamura[5], Naoki Murai[5], Kenji Nakajima[5], A. L. Chernyshev[6], Martin Mourigal[7], Sung-Jin Kim[3] and Je-Geun Park[1,2,*]

[1]Center for Quantum Materials, Seoul National University, Seoul 08826, Republic of Korea

[2]Department of Physics and Astronomy, Seoul National University, Seoul 08826, Republic of Korea

[3]Department of Chemistry and Nano Science, Ewha Womans University, Seoul 03760, Republic of Korea

[5]Materials and Life Science Division, J-PARC Center, Tokai, Ibaraki 319-1195, Japan

[6]Department of Physics and Astronomy, University of California, Irvine, California 92697, USA

[7]School of Physics, Georgia Institute of Technology, Atlanta, Georgia 30332, USA

[$]These authors contributed equally to this work

*jgpark10@snu.ac.kr

The Kitaev model, a honeycomb network of spins with bond-dependent anisotropic interactions, is a rare example of having a quantum spin liquid ground state. Although most Kitaev model candidate materials eventually order magnetically due to inevitable non-Kitaev terms, their bond-dependent anisotropy manifests in unusual spin dynamics. It has recently been suggested that bond-dependent anisotropy can stabilise novel magnetic phases and exotic spin dynamics on the geometrically frustrated triangular lattice. However, few materials have been identified with simultaneous geometric frustration and bond-dependent anisotropy. Here, we report a frustrated triangular lattice with bond-dependent anisotropy in the cobalt-based triangular van der Waals antiferromagnet $CoI_2$. Its momentum and energy-resolved spin dynamics exhibit substantial magnon breakdown and complex level repulsion, as measured by inelastic neutron scattering. A thorough examination of excitations in both the paramagnetic and magnetically ordered states reveals that the bond-dependent anisotropy is the origin of the spiral order and the magnon breakdown found in $CoI_2$. Our result paves the way toward a new research direction for the Kitaev model with geometrical frustration.

Quantum spin liquids are quantum-entangled forms of magnetic matter, first conceptualised by Anderson for a triangular-lattice network of quantum spins[1]. Finding materials that realize Anderson's concept has been challenging. A well-tested route utilises geometrical frustration[2] from non-bipartite lattices such as triangular, Kagome, and pyrochlore spin networks[3]. A recent proposal instead relies on using the bond-dependent anisotropic exchange, i.e., 'exchange frustration'. This seminal work by Kitaev showed that when such interactions decorate a honeycomb structure, the quantum-spin problem is exactly solvable[3]. The resulting quantum spin-liquid hosts both fractional Majorana and gauge field excitations, which have become a central paradigm for an entirely new field of magnetism[3]. Since the discovery of the Kitaev model and its exceptional properties, continuous efforts have been made to realise it in real materials[4]. A promising avenue requires spin-orbital entangled $J_{eff}$=1/2 magnetic moments in an edge-sharing network of ligand octahedrons[5].

While most attempts were made on transition-metal compounds with strong spin-orbit coupling ($4d$/$5d$ shells), recent theoretical work points to $3d$ $Co^{2+}$ as an excellent approach to produce the Kitaev interaction[6,7]. The $3d^7$ $Co^{2+}$ ion has a $^4F$ (i.e. $S = 3$, $L = 2$) multiplet ground state[8,9]. The cubic crystal field (CF) environment splits the multiplet. It yields an $S = 3/2$ and effective angular momentum $L_{eff} = 1$ ground state, which further splits into a spin-orbital entangled $J_{eff} = 1/2$ state when adding spin-orbit coupling. In most known materials, the non-cubic CEF, such as trigonal distortion, further breaks the degeneracy of the multiplet ground states (Fig. 1a)[10]. Indeed, several $Co^{2+}$ honeycomb systems have been since examined in search for dominant Kitaev interactions with the $J_{eff}$=1/2 state of $Co^{2+}$, such as $Na_2Co_2TeO_6$[10–12], $Na_3Co_2SbO_6$[10,11,13], and $BaCo_2(AsO_4)_2$[14,15].

If and how Kitaev physics manifests in geometrically frustrated systems, such as the triangular-lattice antiferromagnets (TLAFs)[16] and pyrochlore-lattice systems, is a much less explored direction and the subject of this work. Only a few $5d$ compounds have been proposed for realizing this physics, for example, triangular-lattice $Ba_3IrTi_2O_9$[17] and pyrochlore-lattice $La_2(Mg/Zn)IrO_6$[18]. However, these materials have no edge-sharing network of ligands, characteristic of Kitaev materials, and are challenging to synthesize in the single-crystalline form for the spectroscopic probes. Hence, finding a true example of Kitaev physics in the geometrically frustrated lattice is an important and urgent issue.

In the case of TLAFs, magnetic excitations are expected to experience strong quantum effects because of the noncollinear magnetic order and significant quantum fluctuations. One such quantum effect is magnon decay, in which magnons break down into multiple magnons or other types of quasiparticles. While this effect usually manifests via a significant broadening of the spectral lines[19,20], in the strong interaction limit, magnons can undergo the avoided decay[21] or be deconfined into spinons[22]. Few direct experimental observations of such spontaneous quasiparticle breakdown have been made in general (such as $Ba_3CoSb_2O_9$, $BiCu_2PO_6$)[23–28] and, given the exotic nature of the $S$=1/2 Kitaev system on the triangular lattice, it is essential to determine the nature of the spin dynamics in that case.

In this work, we report the spin dynamics of the first example of triangular-lattice Kitaev system: $CoI_2$, a member of the van der Waals (vdW) triangular lattice $TMX_2$ (TM = Transition metal, $X$ = Halide) family[29–32]. $CoI_2$ comprises layers of $CoI_6$ octahedrons connected in an edge-sharing network to form a triangular lattice (Fig. 1b), an ideal starting point to search for Kitaev interactions. Our inelastic neutron scattering data show the unmistakable sign of a spin-orbit exciton at 35.4 meV, corresponding to a transition between the $J_{eff}$=3/2 and 1/2 multiplets of the spin-orbit entangled $3d^7$ manifold (See Supplementary Fig. 2). The large gap between these excitations indicates that we can model the $J_{eff}$=1/2 states without considering the higher-$J$ states in the low-energy Hamiltonian. $CoI_2$ orders antiferromagnetically at $T_N$ = 9 K, where the $J_{eff} = 1/2$ moments form a long-period commensurate spiral order in the $ab$ plane and are coupled antiferromagnetically between planes with a propagation vector $\mathbf{Q}_m$ = (1/8, 0, 1/2) (Fig. 1c)[31,33].

To elucidate the nature of exchange interactions in CoI$_2$, we investigated the spin excitations of a single-crystal sample using inelastic neutron scattering. The result demonstrates that bond-dependent nearest-neighbour (n.n) anisotropic exchange (noted $J_{\pm\pm}$ in the following) and 3$^{rd}$ nearest-neighbour interactions are necessary to explain the spin correlations in the paramagnetic regime and the magnetic order below $T_N$. Using a linear spin-wave theory (LSWT) with these exchange interactions, we demonstrate that our spin Hamiltonian can describe the overall shape of the spin-wave spectra below $T_N$. At the same time, magnon decay and renormalisation effects are prevalent over vast regions of the momentum space, indicating quantum effects going beyond LSWT. Our calculations of the two-magnon density of states (DOS) reveal the coexistence of two otherwise distinct possibilities: the conventional, perturbative, magnon decay mechanism via the two-magnon continuum and the magnon survival effect originating from significant level repulsion in the strongly interacting regime.

For microscopic analysis, we start with a magnetic Hamiltonian for CoI$_2$ obtained from symmetry analysis[16,30],

$$H = \sum_{<i,j>_n}^{n=1} \{J_1[S_i^x S_j^x + S_i^y S_j^y + \Delta_1 S_i^z S_j^z] + 2J_{\pm\pm}[(S_i^x S_j^x - S_i^y S_j^y)\cos\phi_\alpha - (S_i^x S_j^y + S_i^y S_j^x)\sin\phi_\alpha]$$
$$+ J_{z\pm}[(S_i^y S_j^z + S_i^y S_j^z)\cos\phi_\alpha - (S_i^x S_j^z + S_i^z S_j^x)\sin\phi_\alpha]\} + \sum_{<i,j>_n}^{n=3} J_3[S_i^x S_j^x + S_i^y S_j^y + \Delta_3 S_i^z S_j^z]$$
$$+ \sum_{<i,j>_n}^{n=2,c_1,c_2} J_n[S_i^x S_j^x + S_i^y S_j^y + S_i^z S_j^z], \quad (1)$$

with the bond-dependent phase factor $\phi_\alpha \in \{0, \frac{2\pi}{3}, -\frac{2\pi}{3}\}$ and each phase factor represents the X-, Y-, and Z-bonds, respectively (Fig. 1b and 1d). In the following, we neglect the allowed $J_{z\pm}$ anisotropy and further assume that 2$^{nd}$ n.n. intra-layer and inter-layer couplings are Heisenberg-type (See Supplementary Notes and Supplementary Fig. 6).

First, we revisit the magnetic phase diagram of this model using the standard Luttinger-Tisza (LT) method[34] with ferromagnetic $J_1 < 0$. We draw the magnetic phase diagram of the spin-isotropic $J_1$-$J_2$-$J_3$ model (See Supplementary Fig. 3); this exhibits a competition among further n.n. interactions that yield a rich magnetic phase diagram consistent with the previous reports.[33] In this model, the propagation vector observed for CoI$_2$ can be stabilised if the following condition can be satisfied: $2\sqrt{2}J_3 + (1+\sqrt{2})J_2 + J_1 = 0$ (See Supplementary Notes). This condition is then found to require a sizable 2$^{nd}$ n.n. interaction $J_2$, comparable to $J_1$. However, this is incompatible with most experimental studies on Co$^{2+}$ edge-sharing compounds that convincingly evidenced small or negligible $J_2$ and strong $J_3$[10–12,14,35]. To resolve this problem, we instead adopt another model with bond-dependent anisotropy and negligible $J_2$, that is, a $J_1$-$J_{\pm\pm}$-$J_3$ model (Fig. 1f). Given that Dzyaloshinskii-Moriya interactions and biquadratic interaction are not allowed in our system because of the space group symmetry and the spin value of $S = 1/2$, this can be considered a minimal model to explain the magnetic structure of CoI$_2$. According to our examination of the magnetic phase diagram, the model without bond-dependent anisotropy ($J_{\pm\pm} = 0$) produces a ground state with $Q_m$ // (h, h, 1/2), inconsistent with the $Q_m$ = (1/8, 0, 1/2) order observed in CoI$_2$. This result clearly shows the central importance of bond-dependent anisotropy in CoI$_2$. In the most general case, the observed $Q_m$ = (1/8, 0. 1/2) constrains the exchange parameters of our model by a single equation $2\sqrt{2}J_3 + (1+\sqrt{2})J_2 + |J_{\pm\pm}| + J_1 - 2J_{c2} = 0$.

To gain more quantitative insight into the exchange constants of CoI$_2$, we use the magnetic scattering measured in the paramagnetic regime, where only short-range correlations are present without magnetic order. This allows us to estimate exchange parameters more reliably because the effect of quantum fluctuations is supposedly small in the paramagnetic state[30,36–38,] and thus the spin dynamics reach a semi-classical regime[30,36–38]. Fig. 2 shows the diffuse scattering of CoI$_2$ taken at $T$ =

13 K in the paramagnetic state, with the energy-integrated diffuse scattering having a hexagonal shape with edges parallel to the Brillouin Zone (BZ) boundaries. Lines then connect the hexagons at the BZ K points. Energy and momentum-resolved cuts along the high-symmetry momentum directions of the BZ reveal highly dispersive paramagnetic scattering signals just above $T_N$ (Fig. 2e), consistent with our expectations of strong magnetic correlations and fluctuations in this system.

The energy-resolved paramagnetic excitations appear to be a stringent test in determining the correct exchange model. Indeed, we simulated the paramagnetic excitations using Landau-Lifshitz dynamics (LLD)[39,40] for two minimal intra-layer coupling models, the $J_1$-$J_2$-$J_3$ model and the $J_1$-$J_{\pm\pm}$-$J_3$ model. For the $J_1$-$J_2$-$J_3$ model, we used $J_1$ = -3.8 meV, $J_2$ = 1.52 meV, $J_3$ = -0.32 meV, $J_{c1}$ = -0.38 meV, $J_{c2}$ = 0.38 meV and $\Delta_1$ = 0.93 as obtained from the bandwidth of the paramagnetic excitation and the constraints from the LT method. For the $J_1$-$J_{\pm\pm}$-$J_3$ model, we instead adapted $J_1$ = -2.73 meV, $J_{\pm\pm}$ = -0.41 meV, $J_3$ = 1.01 meV, $J_{c1}$ = 0.27 meV, $J_{c2}$ = 0.27 meV, and $\Delta_1$ = 0.95. As is shown in Fig.2, the energy-integrated scattering signature differs between the two models, especially in the shape and intensity of the hexagonal patterns. This scattering signature mainly comes from $J_2$ and $J_3$ (See Supplementary Figs. 4 and 5). Moreover, when we compare these two models using the energy-resolved data through cuts along the high-symmetry directions, important differences are observed along the Γ-K direction: a flat-dispersion-like behaviour is only consistent with the $J_1$-$J_{\pm\pm}$-$J_3$ model. This result again demonstrates that the bond-dependent exchange anisotropy is important in CoI$_2$ and is responsible for the observed noncollinear spiral order.

With this knowledge, we now turn to the spin-wave spectra of CoI$_2$ in the magnetically ordered state at $T$ = 4 K. Broad-band inelastic scattering measurements using several incoming neutron energies are necessary to capture all the details of the spectra simultaneously. Fig. 3 summarises the energy-resolved data cut along the high-symmetry directions with two incident neutron energies, $E_i$ = 17 and 8 meV. Along the (H, 0, 0) direction at L = -1/2 (see Γ$_2$ to Γ$_1$ direction in Fig. 3), we can see a gapless excitation originating from the spiral magnetic order with $Q_m$ = (1/8, 0, 1/2). This is consistent with either the $J_1$-$J_2$-$J_3$ model or the $J_1$-$J_{\pm\pm}$-$J_3$ model, given the emergent continuous symmetry of the latter[16]. While strong excitations are present at low energy, the excitation linewidth broadens considerably at the BZ boundaries, especially for the Γ-K direction (see an upper panel of Fig. 3). From the $E_i$ = 8 meV data, we also observe that this broadening changes into a complex magnon band splitting as getting closer to the BZ centre having lower excitation energies.

We now compare the data with LSWT simulations for the exchange interactions obtained in the paramagnetic phase with an anisotropic $J_3$ (See Supplementary Fig. 8 and Supplementary Table 1). Since there are three symmetrically equivalent magnetic domains (see Fig. 1e), we have considered the spectrum from each domain with equal weight. While simulated spin waves, including their bandwidth, are comparable to the data, features deviating from LSWT are apparent. For example, the high-energy magnon branch is considerably broader than anticipated, especially for the magnon modes at the BZ boundary. Also, an additional magnon mode split from the central dispersion appears near 1.5 meV along the (H, H, 0) direction (see the green triangle in Fig. 4b), whose gradual dispersion cannot be captured by the LSWT. Finally, we ascribe these features to significant magnon breakdown, which has not been observed in other Co Kitaev systems yet[12–14].

Theoretically, the $S$ = 1/2 TLAF system is known to have both magnon decay and renormalisation effect via three-boson terms. The three-boson (cubic) terms originate from the coupling between transverse and longitudinal spin fluctuations and can be written as

$$H_3 = \sum_{1-3} V_3^{(1)} b_1^\dagger b_2^\dagger b_3 \delta(k_1 + k_2 - k_3) + H.c. , \quad (2)$$

This three-boson interaction leads to a magnon decaying into two magnons with the kinematic constraints (see Fig 4a). The traditional way to achieve this coupling is a noncollinear magnetic structure[19]. Indeed, a few examples have revealed this mechanism in a real system[23,24,41]. On the other hand, it has been recently suggested that the transverse-to-longitudinal coupling can exist with bond-dependent anisotropy even for collinear magnetic structures, especially for α-RuCl$_3$[42–44]. Interestingly, CoI$_2$ hosts both mechanisms, which may lead to stronger magnon decay distinct from that induced by only one mechanism. Yet the noncollinear spiral order in the CoI$_2$ is stabilised by $J_{\pm\pm}$, and we suggest that bond-dependent anisotropy is a crucial source in this system.

To unambiguously demonstrate that significant magnon breakdown is at play in CoI$_2$, we calculated the two-magnon density-of-states (DOS) based on our best model, as shown in Fig. 4, with the data for a clear comparison. The two-magnon DOS usually gives a fair estimate of the magnon decay rate because it counts the number of spontaneous decay channels fulfilling kinematic constraints[10,19,24,42]. The non-interacting two-magnon DOS can be calculated with

$$D(\mathbf{q}, E) = \frac{1}{N} \sum_{i,j} \sum_{\mathbf{k}} \delta(E - E_{\mathbf{k},i} - E_{\mathbf{q}-\mathbf{k},j}), \quad (3)$$

where $\mathbf{k}$ is a set of $\mathbf{q}$ points on the equally spaced mesh in the 1st Brillouin zone, $E_{\mathbf{k},i}$ is the $i$th magnon's energy dispersion with given momentum $\mathbf{k}$, and $N$ indicates the normalisation factor (Fig 4b). The two-magnon DOS overlaps with the one-magnon branches at the high-energy region, consistent with the observed magnon decay (Fig. 4b-c). We have also performed a linewidth fitting of magnon modes along the Γ-K direction in Fig. 4d (same direction in Fig 4b-c). We have found that the full width at half-maximum (FWHM) of magnon modes is significantly increased when the one-magnon branches enter the calculated two-magnon continuum in the LSWT calculation. Moreover, the FWHM of the decay area is larger than the instrumental resolution (0.3 meV for $E_i$ = 17meV, 0.15 meV for $E_i$ = 8 meV). Such a significant overlap is a distinct feature of CoI$_2$ that can be primarily attributed to the survival of gapless Goldstone modes in the anisotropic model. This is rare since anisotropic exchange interactions usually open an energy gap in other systems[10,45,46].

We further observe that the magnon mode near 1.5 meV exists far below the region with sizable two-magnon DOS (green triangle in Fig. 4c & 4e). However, according to our LSWT simulation, even the lowest-energy magnon branch largely overlaps with the two-magnon DOS as momentum transfer gets closer towards the $K$ point. A most natural interpretation would be then that the lowest magnon branch in the LSWT avoids the two-magnon continuum through significant renormalisation and can show up as the separate branch as indicated by the green triangle in Fig. 4c. Such an avoided crossing would come naturally from a significant repulsion by the strong interaction between quasiparticles[21,23]. Therefore, both decay and avoided decay coexist in the spin dynamics of CoI$_2$.

Our work uncovers extremely rich spin dynamics for the cobalt Kitaev physics on a triangular lattice as realised through the van der Waals noncollinear antiferromagnet CoI$_2$. The bond-dependent exchange anisotropy plays a crucial role in explaining the formation of the noncollinear magnetic order in CoI$_2$. Moreover, we observed substantial magnon decay and selectively avoided decay over the broad momentum space. These phenomena can be understood by strong magnon-magnon interactions triggered by bond-dependent anisotropy and noncollinear magnetic order. Therefore, our finding is essential for understanding the complex interplay between bond-dependent anisotropy and noncollinear magnetic order in quantum magnets.

## Methods

### Sample preparation

Single-crystal of CoI$_2$ was synthesised using the customised Bridgman furnace. 5g of anhydrous CoI$_2$ powder (Alfa Aesar, 99.999% purity) is used as a starting material. The CoI$_2$ powder was sealed in an evacuated silica tube (14 mm diameter, 1 mm wall thickness) under the pressure of ~$10^{-4}$ Torr. The sealed tube was heated to 535 °C for 6 h and then held at 535 °C for 110 h. The tube was pulled down at 535 °C with a growth rate of 0.5 mm*h$^{-1}$. The resulting single crystal was 2-30 mm long and 14 mm in diameter with shiny-black colour.

### Inelastic neutron scattering (INS) measurements

All INS measurement was performed using the cold-neutron disk chopper spectrometer AMATERAS[47] at the J-PARC in Japan. One piece of a 5g single crystal was coated with the hydrogen-free glue CYTOP to prevent any possible degradation from exposure to air. The prepared single crystal was finally mounted on an aluminium holder and sealed with an aluminium can and indium wires to prevent degradation. The crystal was mounted with the *a-b* plane in the horizontal scattering plane. Then, the sample was cooled in a bottom-loading closed-circle refrigerator for measurements at $T$ = 4 K in the ordered phase and $T$ = 13 K in the paramagnetic phase. Using the repetition-rate multiplication (RRM) method, we could simultaneously access three different incident energies of $E_i$ = 52.43, 17.26, and 8.48 meV. Our data were collected with the two configurations: we rotated the sample from -90° to 90° in steps of 0.5° for the 4 K data, while the angular coverage was reduced from -90° to 30° in steps of 0.5° for the 13 K data. To estimate the magnon energy linewidth broadening, we performed magnon peak fitting. We used a Lorentzian function convoluted with a Gaussian function to fit the quasi-elastic peaks.

### LLD and LSWT calculations

Paramagnetic scattering data were theoretically simulated using the LLD method.[39,40] The supercell size for the Monte-Carlo simulation was set for 64 x 64 x 4. Before sampling, the initial spin configuration was thermalised using the standard Metropolis algorithm over 4000 Monte-Carlo steps. After thermalisation, the spin configuration was integrated by the Landau-Lifshitz equation of motion:

$$\frac{d}{dt}\boldsymbol{S}_i = -\boldsymbol{S}_i \times \left[\frac{\partial H}{\partial \boldsymbol{S}_i} + \alpha \left(\boldsymbol{S} \times \frac{\partial H}{\partial \boldsymbol{S}_i}\right)\right]$$

where $\boldsymbol{S}_i$ is the *i*th spin in the actual configuration, $H$ is the magnetic Hamiltonian, and α is a damping parameter. The integration parameters are set at $\Delta t$ = 0.005 meV$^{-1}$ (the step size of the integration), $\tau$ = 62.83 meV$^{-1}$ (the time interval of the evolution for each sampling), and α = 0.2. The Landau-Lifshitz equation was mapped into the mean-field Schrödinger equation for spin dynamics simulation, calculated using the implicit midpoint method[39,40]. The sampled configurations are then Fourier transformed to get the dynamical structure factor S(Q,w). Lastly, our results based on the classical spin approach are rescaled by a factor of $\beta\omega$ because of classical-quantum correspondence in the LSWT framework[38]. The LLD simulation is based on *Su(n)ny*, an open-source code for simulating general spin systems[48].

We performed our spin-wave simulation using our in-house code based on LSWT with Holstein-Primakoff transformation (See Supplementary Notes for detailed calculation). After calculating the spin waves, the dynamical structure factor was convoluted with the instrumental resolution using HORACE[49]. To consider the broadening effect from a finite momentum integration in the data, LSWT simulation cuts along the high-symmetric directions are integrated like the data. Finally, two-magnon DOS was calculated based on LSWT results.

**Acknowledgement**

We thank Pavel Maksimov, Kipton Barros, Matthew Willson, and Cristian D. Batista for fruitful discussions. This work was supported by the Leading Researcher Program of the National Research Foundation of Korea (Grant No. 2020R1A3B2079375). The INS experiment was performed at the MLF of J-PARC under a user program (Proposal No. 2020B0407). The work of A.L.C. was supported by the U.S. Department of Energy, Office of Science, Basic Energy Sciences under Award No. DE-SC0021221. The work of M.M. was supported by the U.S. Department of Energy, Office of Science, Basic Energy Sciences, Materials Sciences and Engineering Division under award DE-SC-0018660. The work of S.K. and S.-J.K was supported by the Pioneer Research Center Program through the National Research Foundation of Korea funded by the Ministry of Science, ICT & Future Planning (NRF-2022M3C1A309198811).

**Author contributions**

J.-G.P. initiated and supervised the project. S.K., C.K., and S.-J.K. grew the single crystal using the Bridgman furnace. C.K. aligned the sample for measurements. S.O.K., N.M., C.K., P.P., T.K., J.-G.P. and K.N. performed the INS measurement. C.K. analysed the data and performed an LSWT calculation. C.K. and M.M. performed the LLD simulations. P.P., M.M. and A.L.C. contributed to the theoretical interpretation and discussion. C.K., M.M., and J.-G.P. wrote the manuscript with input from all co-authors.

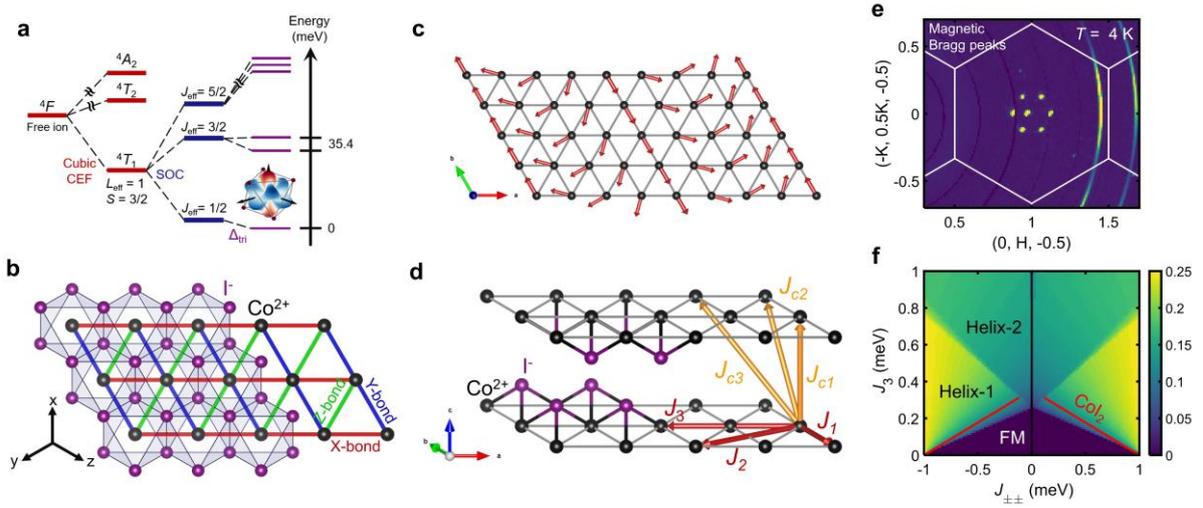

**Fig 1. Schematic view of Kitaev model, spin-orbital entangled $J_{eff}$ = 1/2 state, spiral magnetic structure and magnetic phase diagram of CoI$_2$.**

**a,** Single-ion picture of spin-orbital entangled $J_{eff}$=1/2 state in Co$^{2+}$ ions. Starting from the free ion, the multiplet state of the Co$^{2+}$ ion is split by cubic CEF, spin-orbit coupling (SOC), and trigonal distortion $\Delta_{tri}$. The inset figure shows the visualization of $J_{eff}$=1/2 ground state using the wave function given in Ref[2]. Red indicates up spin, and blue indicates down spin. **b,** Schematic structure of Kitaev model in a triangular lattice. The black and purple spheres indicate Co$^{2+}$ and I$^-$ ions of the CoI$_2$ structure, respectively. Each coloured line connecting Co$^{2+}$ ions represents the 1st *n.n.* Interactions associated with the local *x,y, and z* axes in the Kitaev model. **c,** Magnetic structure of CoI$_2$. The spins rotate on the *ab* plane with a period of 8 unit cells. **d,** The crystal structure of CoI$_2$ and related exchange interaction path. The red arrow indicates intra-layer coupling, and the orange arrows indicate inter-layer coupling. **e,** Magnetic Bragg peaks in CoI$_2$ from elastic neutron scattering data. The six peaks indicate the magnetic Bragg peak of $Q_m$ = (1/8, 0, 1/2) with symmetrically equivalent by 120° rotation. **f,** Magnetic phase diagram of the $J_1$-$J_{\pm\pm}$-$J_3$ model based on the Luttinger-Tisza method. Note that the direction of the propagation vector distinguishes the helix-1 and helix-2 phases. The helix-1 case is parallel to the *a\** direction, whereas the helix-2 phase is parallel to the *a* direction. The colour intensity map represents the length of the propagation vector with r.l.u. The solid red line in each phase diagram indicates the constraints that stabilise the observed magnetic order of CoI$_2$.

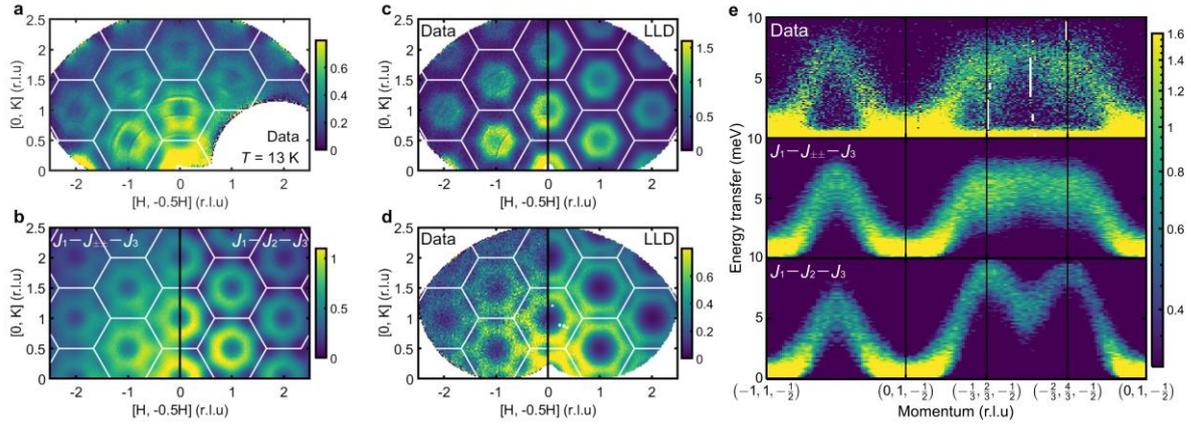

**Fig 2. Diffuse and inelastic scattering in the paramagnetic regime of CoI$_2$**

**a**, Diffuse neutron-scattering intensity measured at $T$ = 13 K by integration of energy transfer E = [1, 7] meV and along L-direction with L = [-0.7, 0.7] r.l.u. White lines indicate the BZ boundaries. **b**, Comparison of the $J_1$-$J_{\pm\pm}$-$J_3$ model and the $J_1$-$J_2$-$J_3$ dominant model using the LLD with the same integration condition taken from data. **c-d**, Comparison of constant-energy slices between experimental data and LLD simulation of $J_1$-$J_{\pm\pm}$-$J_3$ at E = [1, 3] meV and [4, 7] meV for **c,d** respectively. **e,** Energy-momentum slice of the scattering intensity along the high-symmetric points in the Brillouin zone and comparison with the model. The data are measured with the incident energy $E_i$ = 17 meV and integrated with 0.2 Å$^{-1}$ for the L direction and 0.036 Å$^{-1}$ for the in-plane direction. A logarithmic intensity scale was used for better presentation.

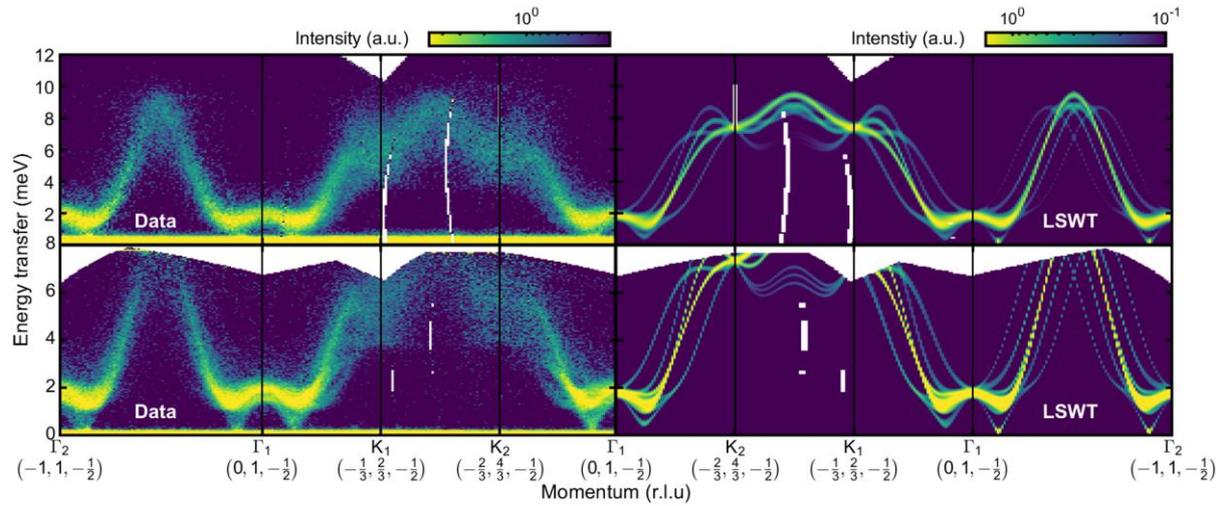

**Fig 3. Spin-wave excitations and magnon decay in magnetically ordered CoI$_2$**

Comparison between energy and momentum-resolved INS cross-section of data (left) and LSWT simulation (right) at $T$ = 4 K. The logarithmic scale was used for a better presentation of data. The upper (lower) row data is measured with the incident energy $E_i$ = 17(8) meV, and the data was integrated with 0.2 Å$^{-1}$ for the $L$ direction and 0.036 Å$^{-1}$ for the in-plane direction. In the LSWT simulation, we convoluted the calculated spectra with the experimental resolution of FWHM = 0.3(0.15) meV for each incident energy. The data and simulation results are plotted with identical momentum integration ranges.

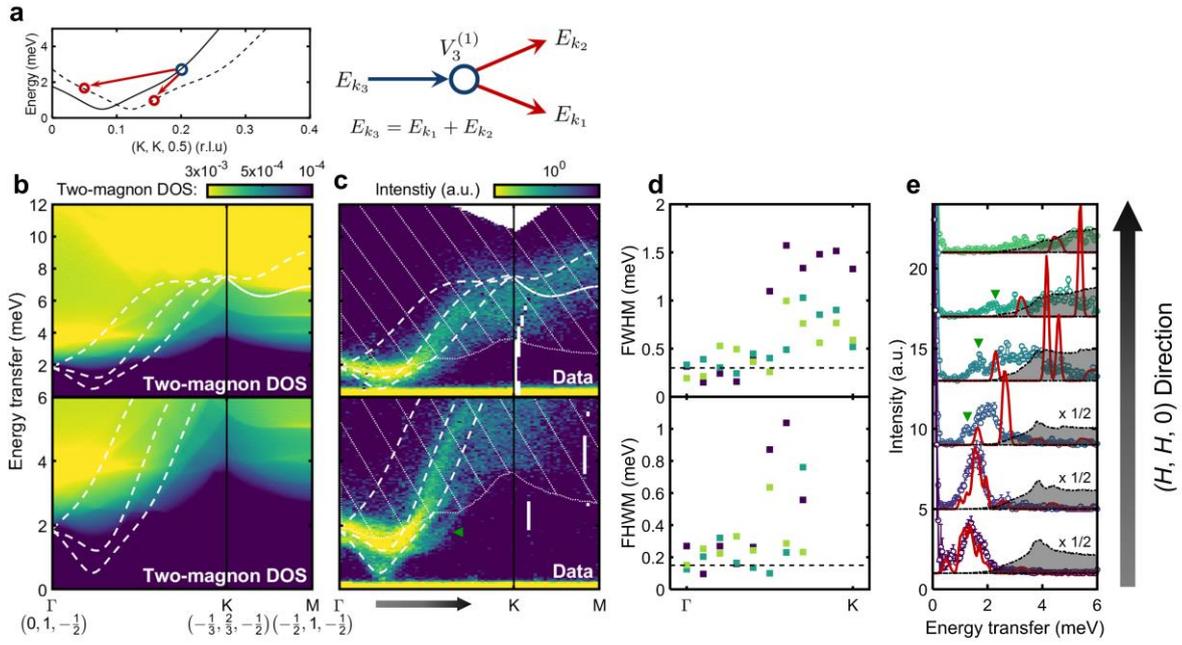

**Fig 4. Two-magnon DOS, magnon decay, and selective avoided decay mode**

**a,** Schematic picture of two-magnon scattering process and three-boson terms. The left figure shows the two-magnon decay of a single magnon with kinematic constraints based on our LSWT calculation. The right figure is a corresponding vertex of the three-boson term. **b,** Two-magnon DOS simulation based on LSWT. The upper (lower) column shows the energy scale with $E$ = 12 (6) meV. White dashed lines indicate the LSWT modes for one magnetic domain, and the colour scale is the two-magnon density of states. **c,** Inelastic neutron scattering data compared to the kinematic extent of the two-magnon continuum. The dotted line represents the boundary of the two-magnon continuum, and the dashed lines represent the LSWT solution. The upper (lower) column data were measured with the incident energy $E_i$ = 17(8) meV. **d,** Fitted FWHM of the magnon modes along the G-K line in Fig. 4**c**. Black dashed lines indicate the instrumental resolution of each incident energy $E_i$ = 17(8) meV. **e,** Constant momentum slices for comparison between the data and LSWT with the incident energy $E_i$ = 8 meV. Each momentum slice was cut along the equal direction, the grey arrow below Fig. 4**c**. Grey shaded area for each slice indicates the two-magnon DOS. The green triangle indicates the avoided decay mode shown in Fig. 4**c**.

13